 \documentclass[preprint2]{aastex} 

\slugcomment{}

\shorttitle{Predicting the solar maximum with the rising rate}
  \shortauthors{Z. L. Du \& H. N. Wang}

\begin{document}
\title{Predicting the solar maximum with the rising rate}

\author{Z. L. Du and H. N. Wang\altaffilmark{}}
\affil{Key Laboratory of Solar Activity, National Astronomical
Observatories, Chinese Academy of Sciences, Beijing 100012, China}
\email{zldu@nao.cas.cn}

\begin{abstract}
The growth rate of solar activity in the early phase of a solar
cycle has been known to be well correlated with the subsequent
amplitude (solar maximum). It provides very useful information for
a new solar cycle as its variation reflects the temporal evolution
of the dynamic process of solar magnetic activities from the
initial phase to the peak phase of the cycle. The correlation
coefficient between the solar maximum ($\beta_{\mathrm{a}}$) and
the rising rate ($\beta_{\mathrm{a}}$) at $\Delta m$ months after
the solar minimum ($R_{\mathrm{min}}$) is studied and shown to
increase as the cycle progresses with an inflection point
($r=0.83$) at about $\Delta m=20$ months. The prediction error of
$R_{\mathrm{max}}$ based on $\beta_{\mathrm{a}}$ is found within
estimation at the 90\% level of confidence and the relative
prediction error will be less than 20\% when $\Delta m\ge20$. From
the above relationship, the current cycle (24) is preliminarily
predicted to peak around October, 2013 with a size of
$R_{\mathrm{max}}=84\pm 33$ at the 90\% level of confidence.
\end{abstract}

\section{Introduction}           
     \label{sect:intr}

The Waldmeier effect that stronger cycles tend to rise faster is a
well known fact in solar activity
\citep{1Waldmeier,2Hathaway,3Du}. The growth rate of solar
activity ($R_{\rm z}$) in the early phase of a solar cycle,
defined as the ratio of a given increment ($\Delta R_{\rm z}=20$)
between two certain levels ($R_{\rm z1}=30$ and $R_{\rm z2}=50$)
over the corresponding elapsed time ($\Delta t$), was found to be
highly correlated ($r>0.8$) with the subsequent amplitude
\citep{4Cameron}. Therefore, the strength of a new cycle should be
rationally predicted by the above relation. The problem is that if
and at what month after the start of a new cycle the strength of
the cycle can be well estimated from the early information of the
cycle.

This paper studies the variation of the correlation between the
maximum amplitude ($R_{\mathrm{max}}$) of a solar cycle and the
rising rate ($\beta_{\rm a}$) as a function of $\Delta m$ months
entering the cycle and analyzes the predictive power of
$\beta_{\mathrm{a}}$ on $R_{\mathrm{max}}$ in order to find out at
what month $R_{\mathrm{max}}$ can be well estimated by $\beta_{\rm
a}$. The results are shown in the following section. $\beta_{\rm
a}$ is defined as the ratio of the increment of $R_{\rm z}$ from
the minimum ($R_{\rm min}$) over the elapsed time ($\Delta m$
months). The temporal variation in the correlation coefficient
($r$) between $R_{\mathrm{max}}$ and $\beta_{\mathrm{a}}$ is
analyzed in Section~\ref{subsec:correlation}, showing that $r$ is
very low near the initial phase ($r<0.5$ if $\Delta m\le10$) and
significant only at a few months after the start of the cycle
($r>0.8$ if $\Delta m\ge19$). The predictive power of
$\beta_{\mathrm{a}}$ on $R_{\mathrm{max}}$ as the cycle progresses
is analyzed in Section~\ref{subsec:power}, indicating that the
relative prediction error of $R_{\mathrm{max}}$ is very small for
almost all $\Delta m$ in some cycles and smaller than 20\% at some
(about twenty) months after the start in other cycles. The peak
size and its timing of cycle 24 are estimated in
Section~\ref{sec:Prediction}, followed by conclusions in
Section~\ref{sec:Discussions}.  

\section{Data and Analysis} \label{sec:Data}

The data used in the present study are the smoothed monthly mean
international sunspot number ($R_{\rm
z}$)\footnote{http://www.ngdc.noaa.gov/stp/spaceweather.html} from
July, 1749 to February, 2011. The rising rate is defined as the
ratio, $\beta_{\rm a}=(R_{\rm z}(\Delta m)-R_{\rm min})/\Delta m$,
of the increment of $R_{\rm z}$ from the minimum ($R_{\rm min}$)
over the elapsed time ($\Delta m$) from the start of the cycle.
The rising rate is computed for each cycle n and each $\Delta m$,
denoted by $\beta_{\rm a}(\Delta m,n)$. The parameters are listed
in Table \ref{Tab:tab1} in which $R_{\mathrm{min}}$ and
$R_{\mathrm{max}}$ are the minimum and maximum amplitudes of the
solar cycle, respectively; $T_{\mathrm{a}}$ is the rise time from
minimum to maximum; $\beta_{\rm a}(27,n)$ is the value of
$\beta_{\rm a}(\Delta m,n)$ at the current state $\Delta m=27$;
other parameters will be described later; and the last row
indicates the relevant averages over cycles $n=7$\,--\,23.

\begin{table}[!tb]
 \small 
 \tabcolsep 0.6mm
 \caption{Parameters and results}
  \label{Tab:tab1}
 \begin{tabular}{crrcccc}
 \tableline  
 $n $& $R_{\mathrm{min}} $& $R_{\mathrm{max}} $&
$T_{\mathrm{a}} $& $\beta_{\rm a}(27,n) $
 &$\overline{E}$ ($\overline{E}_{\mathrm{t}}$)
 &$cc$ ($cc_{\mathrm{t}}$)\\
  \tableline
 1&  8.4&  86.5& 76& 0.74 \\
 2& 11.2& 115.8& 40& 2.23\\
 3&  7.2& 158.5& 35& 3.94\\
 4&  9.5& 141.2& 41& 3.64\\
 5&  3.2&  49.2& 82& 0.45\\
 6&  0.0&  48.7& 70& 0.20\\
 7&  0.1&  71.5& 79& 0.63 &0.18(0.80) &$-0.15(-0.98)$\\
 8&  7.3& 146.9& 40& 3.58 &0.15(0.27) &$-0.90(-0.98)$\\
 9& 10.6& 131.9& 55& 1.24 &0.32(0.31) &$+0.39(-0.88)$\\
10&  3.2&  98.0& 50& 1.54 &0.13(0.47) &$+0.11(-0.98)$\\
11&  5.2& 140.3& 41& 2.49 &0.09(0.32) &$-0.32(-0.98)$\\
12&  2.2&  74.6& 60& 1.76 &0.39(0.57) &$-0.35(-0.98)$\\
13&  5.0&  87.9& 47& 2.34 &0.30(0.49) &$+0.18(-0.98)$\\
14&  2.7&  64.2& 49& 1.42 &0.43(0.67) &$-0.65(-0.96)$\\
15&  1.5& 105.4& 49& 1.94 &0.09(0.41) &$-0.01(-0.93)$\\
16&  5.6&  78.1& 57& 1.93 &0.22(0.52) &$-0.30(-0.95)$\\
17&  3.5& 119.2& 43& 1.99 &0.15(0.35) &$-0.29(-0.96)$\\
18&  7.7& 151.8& 39& 2.73 &0.21(0.28) &$-0.81(-0.98)$\\
19&  3.4& 201.3& 47& 5.26 &0.19(0.20) &$-0.96(-0.95)$\\
20&  9.6& 110.6& 49& 2.42 &0.06(0.38) &$-0.22(-0.97)$\\
21& 12.2& 164.5& 44& 3.14 &0.25(0.26) &$-0.93(-0.98)$\\
22& 12.3& 158.5& 34& 4.64 &0.10(0.31) &$+0.65(-0.98)$\\
23&  8.0& 120.8& 47& 2.21 &0.09(0.35) &$-0.82(-0.98)$\\
24&  1.7& ?84.0& ?59& 1.17\\
 \tableline
av.& 5.9& 119.1& 49& 2.43 &0.20(0.41) &$-0.31(-0.96)$\\
 \tableline 
 \end{tabular}
 \end{table}

\subsection{The variation in the correlation between $R_{\rm max}$ and $\beta_{\rm a}$
  as the cycle progresses } \label{subsec:correlation}

\begin{figure}[!tb]
 \includegraphics[width=0.9\columnwidth]{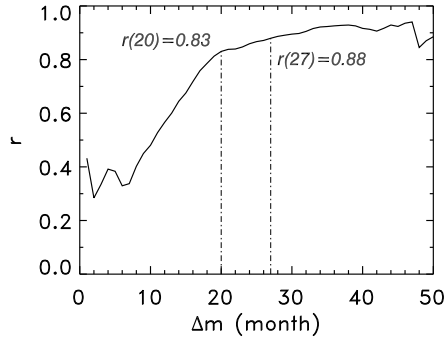}
 \caption{ Correlation
coefficient ($r$) between $R_{\mathrm{max}}$ and
$\beta_{\mathrm{a}}$ as a function of $\Delta m$.
     }
 \label{Fig:1}
 \end{figure}

Figure \ref{Fig:1} illustrates the variation in the correlation
coefficient ($r$) between $R_{\rm max}(n)$ and $\beta_{\rm
a}(\Delta m,n)$ for the cycles in which
$T_{\mathrm{a}}(n)\ge\Delta m$ at a given $\Delta m$ (using only
the data in the rising phases). One can see that $r$ varies with
the progression of the cycle ($\Delta m$). A steady increasing
trend is shown in $r$ since $\Delta m=6$:  $r$ increases from
about 0.33 at $\Delta m=6$ to about 0.83 at $\Delta m=20$ at a
high speed, and increases at a smaller speed since then, showing
an inflection point at $\Delta m=20$ months, $r(20)=0.83$. Near
the initial phase ($\Delta m\le10$) the correlation is not strong
($r<0.5$). The correlation coefficient between $R_{\rm max}$ and
$\beta_{\rm a}$ is high enough at $\Delta m=19$ ($r>0.81$) months
entering the solar cycle. At the current state ($\Delta m=27$),
$r(27)=0.88$.

\subsection{The predictive power of $\beta_{\rm a}$ on $R_{\rm max}$} \label{subsec:power}

\begin{figure*}[!tb]
 \includegraphics[width=2\columnwidth]{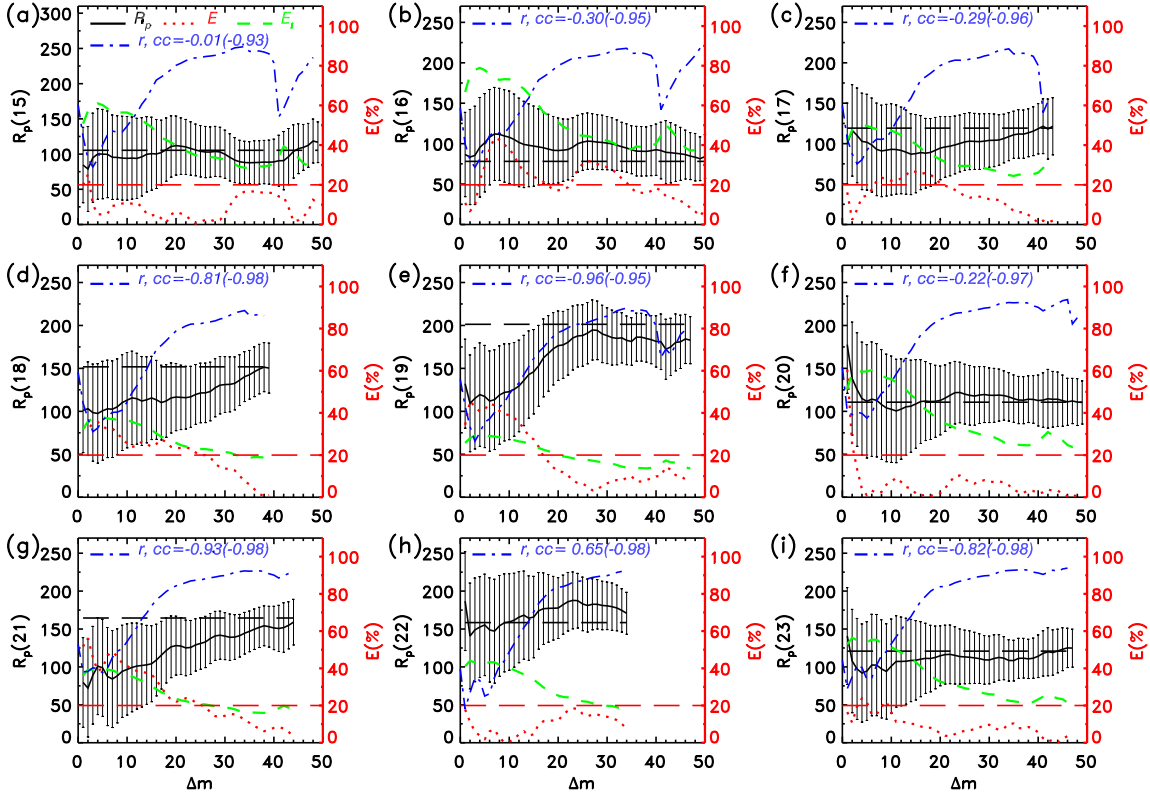}
 \caption{Predictions ($R_{\mathrm{p}}$, black solid line, left
hand scale) of $R_{\mathrm{max}}$ (black horizontal {long-dashed}
line) together with error bars ($t_{\rm r}(n-1)\sigma(\Delta
m,n-1)$, vertical line) for cycles 15\,--\,23 (panels
(a)\,--\,(i), respectively), the actual relative prediction error
($E$, red dotted, right hand scale), the estimated value ($E_{\rm
t}$, green dashed), and the correlation coefficient between
$R_{\mathrm{max}}$ and $\beta_{\rm a}$ ($r$, blue dash-dotted).
The numbers in the figure ($cc$) denote the correlation
coefficients between $E$ ($E_{\rm t}$) and $r$.}
 \label{Fig:2}
 \end{figure*}

In order to test the predictive power of $\beta_{\mathrm{a}}$ on
$R_{\mathrm{max}}$ at different $\Delta m$, we use only the data
up to cycle ($n-1$) to predict $R_{\mathrm{max}}$ for cycle $n$.
For a given $\Delta m$, we calculate the linear regression
equation of $R_{\mathrm{max}}(i)$ against
$\beta_{\mathrm{a}}(\Delta m,i)$ for cycles $i=1,2,\cdots, n-1$ in
the form of
\begin{equation}
  \label{Eq:regressionAB}
   R_{\mathrm{max}} = A + B \beta_{\mathrm{a}},
\end{equation}
and the standard deviation $\sigma(\Delta m,n-1)$ used to estimate
the uncertainty of the prediction of $R_{\mathrm{max}}(n)$. Then,
substituting the value of $\beta_{\mathrm{a}}(\Delta m,n)$ into
this equation, the $R_{\mathrm{max}}$ value for cycle $n$ can be
predicted, which is denoted by $R_{\mathrm{p}}(\Delta m,n)$.
Figure~\ref{Fig:2} shows the results for the recent nine cycles
$n=15$\,--\,23: $R_{\mathrm{p}}(\Delta m,n)$ (black solid line)
together with error bars $t_{\rm r}(n-1)\sigma(\Delta m,n-1)$;
$R_{\mathrm{max}}(n)$ (black horizontal long-dashed line), the
actual relative prediction error (red dotted),
\begin{equation}
  \label{Eq:error}
   E(\Delta m,n) =
   \frac{|R_{\mathrm{p}}(\Delta
   m,n)-R_{\mathrm{max}}(n)|}{R_{\mathrm{max}}(n)};
\end{equation}
the estimated relative prediction error (green dashed),
\begin{equation}
  \label{Eq:error0}
   E_{\rm t}(\Delta m,n) =
   \frac{t_{\rm r}(n-1)\sigma(\Delta m,n-1)}{R_{\mathrm{max}}(n)},
\end{equation}
where $t_{\rm r}(n-1)$ is the $t$-value at the 90\% level of
confidence in a student's t-distribution for $n_{\rm f}=(n-3)$
degrees of freedom; and the correlation coefficient between
$R_{\mathrm{max}}(i)$ and $\beta_{\mathrm{a}}(\Delta m,i)$ ($r$,
blue dash-dotted line, multiplied by 100 to be indicated by the
right hand scale). The numbers in the figure ($cc$) denote the
correlation coefficients between $E$ ($E_{\rm t}$) and $r$.

Figure~\ref{Fig:2}(a)  illustrates the results for cycle 15. It is
seen that $R_{\rm max}$ (black horizontal long-dashed line) is all
within the error bars of $R_{\rm p}$ at the 90\% level of
confidence (vertical lines), $E<E_{\rm t}$, and $E<20\%$ (red
horizontal long-dashed line) when $\Delta m\ge 3$, although there
are some fluctuations in both $E$ and $E_{\rm t}$. The
anti-correlation coefficient between $E_{\rm t}$ and $r$ is very
strong, $cc_{\rm t}=-0.93$, implying that the higher the
correlation coefficient ($r$) between $R_{\rm max}$ and
$\beta_{\mathrm{a}}$, the smaller the estimated relative
prediction error ($E_{\rm t}$) from the extrapolation of the
relationship between $R_{\rm max}$ and $\beta_{\mathrm{a}}$.
However, this is only an estimate in theory rather than in
practice. In fact, the correlation coefficient between $E$ and $r$
is almost zero, $cc=-0.01$, implying that the actual relative
prediction error is almost uncorrelated to $r$. That is to say it
is uncertain whether a more (less) accurate prediction corresponds
to a higher (lower) correlation.

In Figure~\ref{Fig:2}(d)  for cycle 18, we test only the results
for the rising phase $\Delta m\le T_{\rm a}(=39)$. One can see
that both $E$ and $E_{\rm t}$ decrease as $\Delta m$ increases.
The anti-correlation coefficient between $E_{\rm t}$ and $r$ is
also very strong, $cc_{\rm t}=-0.98$. The correlation coefficient
between $E$ and $r$ is highly negative, $cc=-0.81$, implying that
a more (less) accurate prediction can be obtained from a higher
(lower) correlation in this case. In cycle 18, $R_{\rm max}$ is
always within the error bars of $R_{\rm p}$ at the 90\% level of
confidence, $E<E_{\rm t}$. In addition, $E<20\%$ when $\Delta m\ge
26$.

Figure~\ref{Fig:2}(g) shows the results for cycle 21: $\Delta m\le
T_{\rm a}(=44)$. The results are similar to those in
Fig.~\ref{Fig:2}(d): both $E$ and $E_{\rm t}$ decrease as $\Delta
m$ increases; the anti-correlation coefficient between $E$
($E_{\rm t}$) and $r$ is very strong, $cc=-0.93$ ($cc_{\rm
t}=-0.98$). At a small $\Delta m$, $E$ is large: $E>E_{\rm t}$ if
$\Delta m\le 17$. As the cycle progresses, $E$ becomes smaller:
$E<E_{\rm t}$ when $\Delta m\ge 18$; $E<20\%$ when $\Delta m\ge
26$.

The results in other cycles are similar to those above. The main
conclusions in Fig.~\ref{Fig:2} may be summarized as follows.
\begin{enumerate}
\item $R_{\rm max}$ (black horizontal long-dashed) is usually
within the error bars of $R_{\rm p}$ (black solid) at the 90\%
level of confidence (vertical lines), apart from cycles 19 and 21
when $\Delta m\le 16$ and $\Delta m\le 17$, respectively;
  \item the estimated relative prediction error $E_{\rm t}$ (green dashed)
tends to decrease as $\Delta m$ increases; $E_{\rm t}$ is highly
anti-correlated with the correlation coefficient $r$ (blue
dash-dotted), $cc_{\rm t}\approx-0.96$;
  \item the actual relative prediction error $E$ (red dotted)
varies with some fluctuations and tends to decrease as $\Delta m$
increases in some cycles (for $n=16$\,--\,19, 21, 23);
  \item there is no established relationship between $E$ and $r$,
such as $|cc|>0.8$ in cycles $n=18$, 19, 21 and 23, while
$|cc|\le0.3$ in cycles $n=15$, 16, 17 and 20, and $cc$ is even
positive in cycle 22 (0.65);
  \item $E<E_{\rm t}$ for all
$\Delta m$ in cycles $n=15$\,--\,18, 20, 22 and 23; $E<E_{\rm t}$
for $\Delta m\ge 17$ in cycle 19, and for $\Delta m\ge 18$ in
cycle 21;
   \item $E<20\%$ (red {horizontal} long-dashed) since a very few months
entering the cycle ($\Delta m >3$) in cycles 15, 20 and 22;
    $E<20\%$ for $\Delta m\ge 34$ in cycle 16,
    for $\Delta m\ge 21$ in cycle 17,
    for $\Delta m\ge 26$ in cycle 18,
    for $\Delta m\ge 17$ in cycle 19,
    for $\Delta m\ge 26$ in cycle 21;
  and for $\Delta m\ge$ 10 in cycle 23,
 \end{enumerate}

In summary, $\beta_{\mathrm{a}}$ behaved very well in predicting
the subsequent $R_{\mathrm{max}}$: (i) the actual prediction error
(known only when the cycle is over) is usually within estimation
since about twenty months entering the cycle,
$|R_{\mathrm{p}}(\Delta m,n)-R_{\mathrm{max}}(n)|<t_{\rm
r}(n-1)\sigma(\Delta m,n-1)$; (ii) the relative prediction error
is usually less than 20\% since about twenty months into the
cycle; and (iii) $E$ tends to decrease as the cycle progresses. In
some cycles ($n=15$, 20 and 22, see Figures \ref{Fig:2}(a), (f)
and (h)), the relative prediction error is very small ($E< 10\%$)
at a small $\Delta m$ even if the correlation coefficient is low
($r<0.5$). Similar conclusions can also be obtained in other
cycles (not shown): $E<E_{\rm t}$ at about $\Delta m\ge20$;
$E_{\rm t}$ is highly anti-correlated with $r$ ($cc_{\rm t})$;
while there is no established relationship between $E$ and $r$
($cc$). The results of $\overline{E}$ ($\overline{E}_{\rm t}$) and
$cc$ ($cc_{\rm t}$) in cycles 7-23 are shown in Table
\ref{Tab:tab1}, and the relevant averages are indicated by the
last row: $<\overline{E}>=0.20$ ($<\overline{E}_{\rm t}>=0.41$)
and $<cc>=-0.31$ ($<cc_{\rm t}>=-0.96$), where $\overline{E}$ is
the average over $\Delta m$ in a solar cycle and $<\overline{E}>$
represents the average over cycles 7-23. Therefore, a higher
(lower) correlation coefficient does not necessarily yield a more
(less) accurate prediction \citep{5Du,6Du,7Du}.

\section{Prediction $R_{\rm max}$ for Cycle 24} \label{sec:Prediction}

\begin{figure}[!tb]
 \includegraphics[width=0.9\columnwidth]{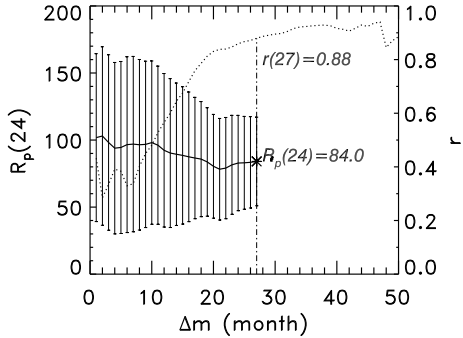}
 \caption{Prediction of
$R_{\mathrm{max}}(24)$ as a function of $\Delta m$.}
 \label{Fig:3}
 \end{figure}

Now, we employ the above technique to predict the peak size of
cycle $n=24$. The results are shown in Fig.~\ref{Fig:3}: $R_{\rm
p}$ (solid) is the predicted $R_{\rm max}(24)$ and $r$ (dotted) is
the correlation coefficient between $R_{\rm max}(i)$ and
$\beta_{\rm a}(\Delta m,i)$ for cycles $i=1,2,\cdots,23$ at a
given $\Delta m$. It is seen that $R_{\rm p}$ does not vary
significantly with $\Delta m$. At the current state ($\Delta
m=27$), the correlation coefficient between $R_{\rm max}$ and
$\beta_{\rm a}$ is $r(27)=0.88$, and the regression equation of
$R_{\mathrm{max}}$ against $\beta_{\mathrm{a}}$ is
\begin{equation}
  \label{Eq:regression24}
   R_{\mathrm{max}} =   52.1+ 27.2 \beta_{\mathrm{a}},
\end{equation}
with a standard deviation of $\sigma=19.2$. Substituting the
current value of $\beta_{\rm a}(27,24)= 1.17$ into this equation,
the peak sunspot number for the ongoing cycle (24) is predicted as
$R_{\mathrm{p}}(24)=84\pm t_{\rm r}(23)\sigma=(84\pm 33$)
(asterisk), where $t_{\rm r}(23)=1.721$ is the t-value at the 90\%
level of confidence in a student's t-distribution for $n_{\rm
f}=23-2=21$ degrees of freedom.

From the relationship between $T_{\mathrm{a}}$ and
$R_{\mathrm{max}}$,
\begin{equation}
  \label{Eq:Ta}
   T_{\mathrm{a}} = 79.7 - 0.251 R_{\mathrm{max}},\quad
   \sigma=9.3,
\end{equation}
one can estimate the rise time $T_{\mathrm{a}}$ for cycle 24.
Using the predicted value (84) of $R_{\mathrm{max}}(24)$, one
obtains $T_{\mathrm{a}}(24)=(59\pm9$) months. Therefore, the peak
of cycle 24 may probably occur around October, 2013, slightly
later than that (May, 2013) by both NASA Marshall Space Flight
Center
(MSFC)\footnote{http://solarscience.msfc.nasa.gov/SunspotCycle.shtml}
based on a quasi-Planck function and NOAA space weather prediction
center
(SWPC)\footnote{http://www.swpc.noaa.gov/SolarCycle/index.html}
based on a consensus decision of ``The Solar Cycle 24 Prediction
Panel¡±, and that (June, 2013) based on a modified Gaussian
function \citep{8Du}.

 \section{Discussions and Conclusions}
\label{sec:Discussions}

Studying the correlation between $R_{\mathrm{max}}$ of a solar
cycle and a related parameter is useful to understand the dynamic
process of the cycle. A high correlation can be used to estimate
the strength of a new solar cycle
\citep{9Kane,10Pesnell,11Messerotti,12Du,13Wang,14Wang,15Wang,16Le,17Li}.
For example, Ohl's geomagnetic precursor method \citep{18Ohl}
succeeded in predicting $R_{\mathrm{max}}$ in cycles 20\,--\,22
\citep{19Layden,20Thompson,21Shastri,22Schussler} due to the high
correlation coefficients ($>0.8$) between $R_{\mathrm{max}}$ and
geomagnetic-based parameters. However, a high correlation does not
always yield a satisfactory prediction
\citep{5Du,6Du,7Du,23Cameron,24Du,25Du} and a low correlation may
also yield an accurate prediction in some cases (see Section
\ref{subsec:power}).

A prominent feature in the solar cycle is the so-called Waldmeier
effect that stronger cycles tend to rise faster
\citep{1Waldmeier,2Hathaway,3Du,4Cameron}. This effect has already
begun to work in the early phase of the cycle
($\beta_{\mathrm{a}}$). The variation in $\beta_{\mathrm{a}}$
reflects the temporal evolution of the dynamic process of solar
magnetic activities from the initial phase to the peak phase of
the cycle, and so, $\beta_{\mathrm{a}}$ can provide very useful
information for the cycle.

In this study, we analyzed the temporal variation in the
correlation coefficient ($r$) between $R_{\mathrm{max}}$ and
$\beta_{\mathrm{a}}$ as a function of $\Delta m$ months after the
solar minimum ($R_{\mathrm{min}}$) and the predictive power of
$\beta_{\mathrm{a}}$ on $R_{\mathrm{max}}$ as the solar cycle
progresses. First, it is shown that $r$ increases as $\Delta m$
increases with an inflection point over 0.8 at about $\Delta m=20$
months. The dynamic process of the solar activity is more
non-linear near the initial phase of the cycle ($r<0.5$ if $\Delta
m\le10$) and tends to be stable after twenty months entering the
cycle.

Besides, $\beta_{\mathrm{a}}$ behaved rather well in predicting
$R_{\mathrm{max}}$: the prediction error is usually within the
estimated one after about $\Delta m=20$ months entering a solar
cycle, $|R_{\mathrm{p}}(\Delta m,n)-R_{\mathrm{max}}(n)|<t_{\rm
r}(n-1)\sigma(\Delta m,n-1)$ at the 90\% level of confidence. This
is a crucial point in prediction because a method will be less
useful if the prediction is not within the prediction range
derived from the method. In addition, the relative prediction
error ($E$) based on $\beta_{\mathrm{a}}$ is usually less than
20\% when $\Delta m\ge20$ months. Thus, $\beta_{\mathrm{a}}$ is a
good indicator for the subsequent $R_{\mathrm{max}}$. Finally, $E$
tends to decrease as the cycle progresses. Therefore, the maximum
amplitude of a new cycle ($R_{\mathrm{max}}$) can be well
estimated at twenty months after the start.

It should be noted in Fig.~\ref{Fig:2} that the correlation
between $R_{\mathrm{max}}$ and $\beta_{\mathrm{a}}$ is not strong
near the initial phase of the cycle, while the prediction of
$R_{\mathrm{max}}$ from $\beta_{\mathrm{a}}$ is rather good in
some cycles (15, 20, 22 and 23). Therefore, a high correlation is
not the sole condition to obtain a more accurate prediction
\citep{23Cameron,24Du,25Du,26Svalgaard,27Schatten}, which may
depend on the variation of the correlation or long-term
periodicities \citep{5Du,6Du,7Du,24Du,25Du}. In this
study,$R_{\mathrm{max}}$ can be well estimated from
$\beta_{\mathrm{a}}$ even if the correlation coefficient ($r$) is
not strong near the initial phase in some cycles, {\it e.g.},
$E<20\%$ for small $r$ ($<0.5)$ at small $\Delta m$ in cycles 15,
20, 22 and 23 (see Fig.~\ref{Fig:2}).

Based on $\beta_{\mathrm{a}}$ at the current state $\Delta m=27$,
the peak sunspot number of the ongoing cycle (24) is predicted to
be $R_{\mathrm{max}}=84\pm 33$ at the 90\% level of confidence or
$R_{\mathrm{max}}=84\pm 19$ with a 1-$\sigma$ uncertainty. This
prediction is higher than a few predictions and lower than many
others, some of which are shown in Table \ref{Tab:tab2}.

\begin{table}[!tb]
 \small 
 \scriptsize
 \tabcolsep 1.mm
 \caption{Some predictions for cycle 24}
  \label{Tab:tab2}
 \begin{tabular}{lrll}
 \tableline  
 Author&
\multicolumn{2}{c}{$R_{\mathrm{max}}\pm\sigma$}&
method or predictor\\
  \tableline
 \citet{9Kane}                        &  58& $\pm25$  & $aa$ minimum\\
 \citet{30Choudhuri}                 &  68&          & solar dynamo model\\
 NASA/MSFC                           &  70&          & quasi-Planck function\\
 \citet{8Du}                           &  72& $\pm11$  & modified Gaussian function\\
 \citet{26Svalgaard}            &  75& $\pm8$   & polar field\\
 \citet{27Schatten}                   &  80& $\pm30$  & polar field\\
 \citet{33Li}                   &  80& /137     & slow/fast riser\\
 \citet{12Du}                          &  82& /53  & $aa$ minimum/corrected\\
 \citet{34Du}                  &  84&$\pm17$   & similar cycles\\
 current study                       &  84&$\pm19$   & rising rate\\
 \citet{31Jiang}                &  85&          & solar dynamo model\\
 \citet{28Li}                   &  88&          & sunspot minimum\\
 NOAA/SWPC                           &  90&          & consensus\\
 \citet{35Wang}            &  97&$\pm25$   & open flux\\
 \citet{15Wang}                 & 100&$\pm8$    & similar cycles\\
 \citet{14Wang}                 & 101&$\pm18$   & similar cycles\\
 \citet{36Hiremath}                    & 110&$\pm11$   & autoregression\\
 \citet{37Rigozo}               & 113&          & spectral components\\
 \citet{38Dabas}                & 124&$\pm23$   & geomagnetic disturbed days\\
 \citet{39Tlatov}                      & 135&$\pm12$   & $G\propto 1/R_{\mathrm{z}}$\\
 \citet{24Du}                   & 140&$\pm16$   & cycle length\\
 \citet{40Hathaway}          & 160&$\pm25$   & I component of $aa$\\
 \citet{32Dikpati}              & 165&$\pm15$   & flux-transport dynamo model\\
 \tableline 
 \end{tabular}
 \end{table}

Accurately predicting the strength of an upcoming solar cycle is
important for both solar physics and solar-terrestrial
environment. A reliable prediction of $R_{\mathrm{max}}$ may test
models for explaining the solar cycle \citep{10Pesnell}. So far, a
great many results have been published on the prediction of
$R_{\mathrm{max}}$ for cycle 24, of which some are based on
statistics and some others are related to physics (see Table
\ref{Tab:tab2}). As the solar activity near the minimum between
cycles 23 and 24 lasts so long a time at a low level before rising
\citep[as shown in the most spotless days since cycle
16,][]{28Li,29Li}, cycle 24 is unusual, which is drawing greater
attention than ever. Besides, as solar dynamo models have begun to
be applied in predicting $R_{\mathrm{max}}$
\citep{30Choudhuri,31Jiang,32Dikpati}, the predictions of the
strength of cycle 24 attract special attention in order to test
the predictive skill of solar dynamo models.

Discrepancies are found in the predictions of $R_{\mathrm{max}}$
for cycle 24 by erent methods 
(Table \ref{Tab:tab2}). Our prediction (84) is near to those by
the polar field (or solar dynamo model), about 30\% lower than the
peak size of cycle 23. Recently, we find that cycle 24 is most
likely similar to cycles 14 and 10 \citep{34Du}. Therefore, even
if cycle 24 is not a strong cycle, large eruption events may also
occur as in cycle 10 for the largest solar storm of the year 1859
(Carrington Event).

Conclusions are summarized below.
\begin{enumerate}
 \item The correlation coefficient ($r$) between the maximum
amplitude ($R_{\mathrm{max}}$) of a solar cycle and the rising
rate ($\beta_{\mathrm{a}}$) at $\Delta m$ months after the solar
minimum ($R_{\mathrm{min}}$) increases as $\Delta m$ increases
with an inflection point at about twenty months entering the
cycle.
  \item The prediction error based on the linear
relationship between $R_{\mathrm{max}}$ and $\beta_{\mathrm{a}}$
is usually within the estimated one when $\Delta m\ge20$,
$|R_{\mathrm{p}}(\Delta m,n)-R_{\mathrm{max}}(n)|<t_{\rm
r}(n-1)\sigma(\Delta m,n-1)$, where $\sigma(\Delta m,n-1)$ is the
standard deviation of the regression equation for the data up to
cycle $n-1$, and $t_{\rm r}(n-1)$ is the $t$-value at the 90\%
level of confidence in a student's t-distribution.
 \item The relative prediction error ($E$) from the above
 technique tends to decrease as the cycle progresses and will be less than 20\% when
$\Delta m\ge20$.
 \item  The
current cycle (24) is temporarily predicted to peak around October
2013 with a size of $R_{\mathrm{max}}=84\pm 33$ at the 90\% level
of confidence.
\end{enumerate}

\section*{Acknowledgments}
The authors are grateful to the two anonymous referees for
suggestive and helpful comments, which
improved the original version of the manuscript. %
This work is supported by the National Natural Science Foundation
of China (Grant Nos. 10973020, 40890161 and 10921303), the
National Basic Research Program of China (grant No. 2011CB811406),
and the China Meteorological Administration (grant No.
GYHY201106011). %


\end{document}